\documentclass[12pt,letterpaper]{article}                  
\usepackage{subfigure}
\usepackage{graphicx}

\begin{document}

\title{Dispersion managed mode-locking dynamics in a Ti:Sapphire laser}

\author{Marco V. Tognetti and Helder M. Crespo\\
\\
CLOQ/Departamento de F\'isica, \\
Faculdade de Ci$\hat{e}$ncias, Universidade do Porto,\\
 Rua do Campo Alegre 687, 4169-007 Porto, Portugal}

\maketitle
\begin{abstract}
We present what is to our knowledge the most complete 1-D numerical analysis of the evolution and the propagation dynamics of an ultrashort laser pulse in a Ti:Sapphire laser oscillator. This study confirms the
dispersion managed model of mode-locking, and emphasizes the role of the Kerr nonlinearity in generating mode-locked spectra with a smooth and $well-behaved$ spectral phase. A very good agreement with preliminary experimental measurements is found.
\end{abstract}

\bigskip

Pulse generation in mode-locked lasers has been a subject of intense research over more 
than two decades. In more recent descriptions of Kerr-lens mode-locked (KLM) ultrafast lasers, pulse formation 
is assumed to rely upon a soliton-like mechanism, based on the master
equation approximation\,\cite{Haus2000-5}, where the combined and balanced action of self-phase modulation (SPM)
and group-delay dispersion (GDD) is the basis for the generation of a short pulse. In this picture the shortest pulse durations are
obtained for cavity configurations with minimum net GDD. This approach gives a good description of actual
systems but does not take into account the pulse propagation dynamics in the single optical elements that comprise
the laser cavity and usually does not include higher order dispersion terms, which are known to significantly  affect 
sub$-10\,fs$ pulses such as those generated with today's state-of-the-art laser oscillators\,\cite{Morgner1999-5,Fuji2003-5}. As an alternative to the master equation approach, the evolution of laser pulses has also been modeled numerically
\cite{Christov1994-5,Christov1996-5}\,, even if the actual sequence of intracavity components was not fully taken into consideration. In fact, the actual ordering of the optical elements is behind a new model for the generation of ultrashort pulses later introduced by Chen
$et\,al.$\,\cite{Chen1999}, on its hand directly related to the nonlinear
propagation of pulses in dispersion managed communication fibers.
They identify a solid state mode-locked laser as another example
of a system where dispersion managed solitons (DMS) can be
observed. DMS are stable soliton-like solutions of
the nonlinear Schr\"odinger equation which are known to occur in
optical media with a periodical change of
sign in the GDD\,\cite{Haus1999}, such as a femtosecond laser cavity. The
main difference from $standard$ soliton propagation is that the
spectrum and temporal profile of DMS periodically broaden and
recompress as the pulse crosses regions of opposite GDD. 

Here we present what is to our knowledge
the most detailed 1-D numerical simulation of a prism-dispersion controlled linear laser cavity, which includes
build up of the laser pulse from noise through the
action of active- and Kerr-lens mode-locking, the measured
reflectivity and phase distortion of every optical element in the cavity, the
measured gain bandwidth of the Ti:sapphire crystal, and the propagation inside the active medium, by numerically solving the corresponding nonlinear Schr\"odinger equation in the presence of gain. In particular, great attention is devoted
to intracavity pulse formation and propagation, showing how the
spectrum and its phase
 evolve as the pulse crosses the crystal, is reflected off the intracavity
 mirrors, and goes through the negatively dispersive prism line.
 This study confirms the dispersion managed model
of mode-locking, showing that a spectrum extending from $700$ to
$950$\,nm with a smooth and {\it nearly flat} phase can be
obtained using commercially available ultrafast optical elements,
in agreement with recent experimental work\,\cite{Crespo2005}.

\begin{figure}[h]
 \centerline{\scalebox{0.5}{\includegraphics[angle=270]{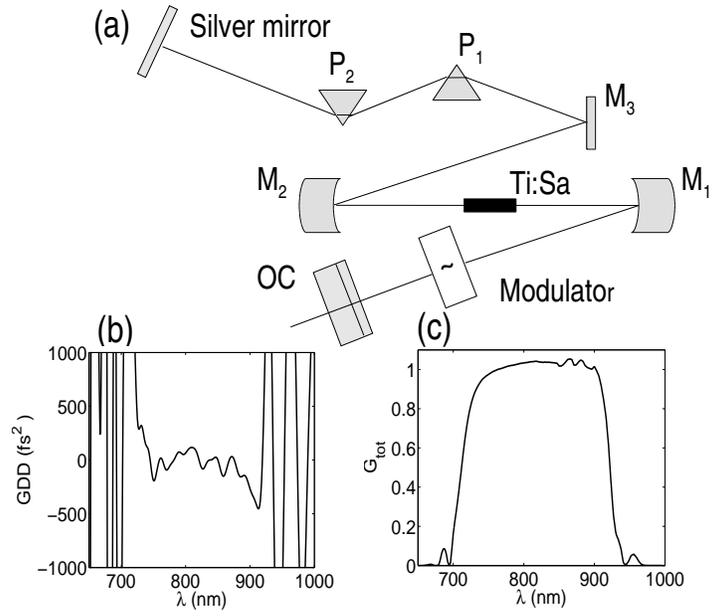}}}
\caption{(a) Schematic diagram of the Ti:sapphire laser oscillator.
(b) Total round trip intracavity GDD. (c) Total cavity gain
$G_{tot}$ assuming a peak crystal gain of 1.04.}
 \label{figure_1_tognetti_crespo}
\end{figure}

\begin{figure}[h]
 \centerline{\scalebox{0.5}{\includegraphics{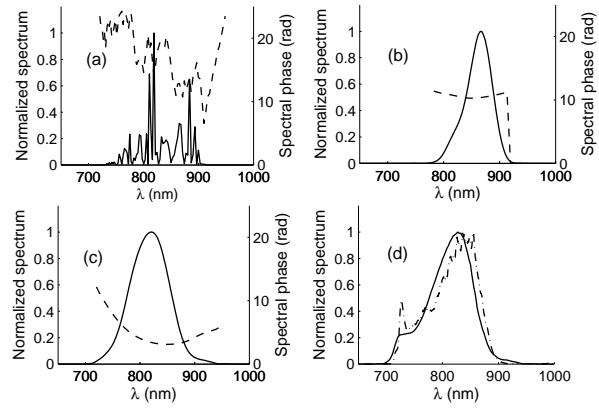}}}
\caption{Normalized intracavity spectrum (solid line) and phase (dotted line) before the OC after
$(a)\,20$, $(b)\,500$, $(c)\,1000$ round trips inside the cavity. (d)
Simulated (solid line) and measured (dashed line) final spectra
outside the OC.}
 \label{figure_2_tognetti_crespo}
\end{figure}

Figure\,\ref{figure_1_tognetti_crespo}\,(a) shows a schematic diagram of the
Ti:Sapphire laser oscillator, which consists of: an active crystal of length $L$
enclosed between two focusing mirrors $M_1$ and $M_2$, a flat folding
mirror $M_3$, an output coupler $OC$ and a silver high reflector at the
cavity ends,
 a pair of fused-silica prisms for dispersion compensation, and an active amplitude modulator $M(t)$
to form the initial pulse from noise. The evolution of the pulse inside the
cavity is described by the following iterative procedure which connects the spectral amplitude of the field,
$\tilde{A}_{k+1}(\omega)$, for the $(k+1)-th$ passage inside the cavity,
with  the field envelope $A_k(t)$, obtained from the
previous passage:
\begin{eqnarray}
\tilde{A}_{k_1}(\omega) &=& R_{1}(\omega)^{\frac{1}{2}}e^{i
\phi_{1}(\omega)}\int_{-\infty}^{+\infty} M(t)A_{k}(t)e^{-i \omega t}\nonumber\\
\tilde{A}_{k_2}(\omega) &=& {\cal P}(\tilde{A}_{k_1}(\omega))\nonumber\\
\tilde{A}_{k_3}(\omega) &=& R_{2}(\omega)R_{3}(\omega)e^{2i(\phi_{2}(\omega)+\phi_{3}(\omega)+\phi_{pr}(\omega))}\tilde{A}_{k_2}(\omega)\nonumber\\
\tilde{A}_{k_4}(\omega) &=& {\cal P}(\tilde{A}_{k_3}(\omega))\nonumber\\
\tilde{A}_{k+1}(\omega) &=& (R_{1}(\omega)R_{OC}(\omega))^{\frac{1}{2}} e^{i (\phi_{1}(\omega)+\phi_{OC}(\omega))} \tilde{A}_{k_4}(\omega),\\
\label{eq_it}\nonumber
\end{eqnarray}
where $R_{i}$ and $\phi_i$ are the reflectivity and phase of the $i-th$
optical element respectively, $M(t)$ is an initial active modulation used to start mode-locked operation \cite{Christov1994-5}, and
$\tilde{A}_{k_{n+1}}(\omega) = {\cal P}(\tilde{A}_{k_n}(\omega))$ is the
spectral amplitude at the crystal output, obtained by numerically solving the following propagation equation inside the crystal:
\begin{eqnarray}
\frac{\partial A(z,t)}{\partial z} &=& \int_{-\infty}^{+\infty} (i
\beta(\omega)+g(\omega)) \tilde{A}(z,\omega)e^{-i\omega t} d
\omega\nonumber\\
& &+i \gamma |A(z,t)|^2 A(z,t) \label{eq_prop}
\end{eqnarray}
with  $A(0,t)=A_{k_n}(t)$. Here
$\beta(\omega)$ and $g(\omega)$ are the crystal phase distortion and
gain profile per unit length, and $\gamma=2.2 \times 10^{-6} W^{-1}\,cm^{-1}$ is the estimated nonlinearity
coefficient. Moreover, self-amplitude modulation (SAM) induced by the Kerr nonlinearity is included as a nonlinear intensity discriminator in the time domain\cite{Christov1994-5}, here modeled as a supergaussian: $K(A)=exp[-\frac{1}{2}(\frac{|A|^2}{\sigma P_0})^m]$, with
$P_0=max(|A|^2)$,  $m=24$ and $\sigma=0.47$. These parameters were determined by trial-and-error, even though the final spectrum will not significantly depend on the exact values, provided that mode-locking operation is established. The crystal gain and the mirror reflectivities and phase
distortions were directly obtained from measurements performed on actual optical components, which comprise a relatively
thick ($4.5\,mm$) Ti:Sapphire crystal (Crystal Systems Inc.), commercially
available standard ultrafast laser mirrors designed for
$850\,nm$ (Spectra-Physics and TecOptics) and a
$3.5\%$ output-coupler (Spectra-Physics).
Figures\,\ref{figure_1_tognetti_crespo}\,(b) and (c) show the total GDD and the total cavity
 gain $G_{tot}$ for one round trip defined as
$G_{tot}(\lambda)=R_1(\lambda)^2R_2(\lambda)^2R_3(\lambda)^2R_{OC}(\lambda)G(\lambda)^2$, with $G(\lambda)=e^{g(\lambda)L}$ the total crystal gain normalized at $1.04$. As initial condition, a $low$ amplitude random noise with flat spectral phase is assumed. The evolution of the intracavity spectral profile and phase of the
pulse prior to entering the output coupler are shown in
figure\,\ref{figure_2_tognetti_crespo}, after \,(a)\,$20$,\,(b)\,$500$,\, and
(c)\,$1000$ round-trips of the pulse inside the cavity. After the
first $20$ round-trips, the spectral phase of the intracavity radiation shows strong fluctuations due to the accumulated linear phase distortion of the optics, as expected from figure\,\ref{figure_1_tognetti_crespo}\,(b). Then a spectrum with
increasingly $smooth$ spectral phase centered around $850\,nm$
builds up from noise mainly through the action of the crystal gain, the
active modulation and SAM, resulting in the spectrum shown in  figure\,\ref{figure_2_tognetti_crespo}\,(b) at a pulse peak power of $3.5\times10^5\,W$. As the pulse intensity increases, new mode-locked frequencies are
generated via SPM and the spectrum is further broadened until a
steady-state is reached at a pulse peak power of $5 \times 10^6\,W$ (see
figure\,\ref{figure_2_tognetti_crespo}\,(c)), as a result of the interplay between
SPM, phase distortion and the finite bandwidth of the total cavity
gain (see figure\,\ref{figure_1_tognetti_crespo}\,(c)).
To illustrate the validity of the simulation code the measured and
the simulated spectra obtained after the OC are given in
figure\,\ref{figure_2_tognetti_crespo}\,(d), revealing a very good agreement
between experimental measurements and theoretical predictions.
The steady-state spectrum of figure\,\ref{figure_2_tognetti_crespo}\,(c) is
reproduced every round-trip but it has a different spectral
profile in different  dispersion regions of the cavity.
Figure\,\ref{figure_3_tognetti_crespo} shows how the cavity can be put in
analogy with a dispersion managed fiber made of a region of
nonlinear propagation and positive GDD (the active medium  crossed
two times), and a region of linear propagation and negative GDD
(the prism-pair crossed two times). The OC and the silver mirror
are placed in the middle point of each dispersion region. In
correspondence with the points $A,B,C,D,E$ and $F$ of
figure\,\ref{figure_3_tognetti_crespo}, figure\,\ref{figure_4_tognetti_crespo} shows how the steady-state pulse
spectrum, phase, and temporal profile evolve when
crossing the crystal, going through the dispersion
compensating prism-pair, and coming back to the OC. 
As the pulse propagates towards the silver mirror, it
enters the active medium with a positive chirp (point $A$ in figures\,\ref{figure_3_tognetti_crespo} 
and\,\ref{figure_4_tognetti_crespo}) and  its spectrum broadens while its temporal width increases due 
to the concomitant action of positive GDD and SPM (points $B$ and $C$). It can be observed 
that spectral broadening mainly takes place in the first half of the crystal (from  $A$ to $B$) 
since temporal broadening due to GDD decreases the strength of SPM as the pulse penetrates more into the crystal.
When the pulse crosses the two prisms and  reaches the silver mirror (point $D$), 
the acquired positive chirp is partially compensated for, while the spectrum remains unaffected,
resulting in a shorter and more intense pulse. This is the point of maximum intracavity spectral width, 
minimum pulse duration and maximum peak intensity. In contrast to the case of point $A$, when the pulse enters 
the crystal on its way back to the OC (point $E$), it is negatively chirped by a second passage in the prism 
sequence: nonlinear propagation  now results in  spectral narrowing and phase flattening until the pulse is 
transform limited at a depth of 3 mm within the crystal (point $F$). This is the point of minimum spectral 
width. As the pulse continues its way to the OC and goes back to the silver mirror, its spectrum is broadened, 
recovering its maximum width at points $C$ and $D$. The pulse behaviour described here confirms the dispersion 
managed model of mode-locking\,\cite{Chen1999}, by showing the typical spectral and temporal $breathing$ of DMS, and also explains the experimental fact that spectra are broader when taken in the dispersive end of KLM lasers\,\cite{asaki1993}. 
Furthermore figure\,\ref{figure_4_tognetti_crespo} shows that the pulse propagating in the cavity, while varying its spectrum
and temporal profile, mantains a $smooth$ spectral phase due to the joint action of SAM and SPM, which are able
to wash out  from the spectral phase the modulations inherent in the net intracavity GDD. This is in agreement
with recent experimental results, where it was shown that operation of a broadband Ti:Sapphire laser under strong Kerr-lens 
mode-locking conditions resulted in a smoothing of the spectral phase\,\cite{Morgner1999-5,Fuji2003-5}.
Once extracted the pulse from the cavity, such a $well-behaved$ phase appears to be suitable for further 
extracavity pulse compression, giving for the spectrum of point $D$ a Fourier transform limited pulse duration of $11\,fs$, also in good agreement with recent experimental results\,\cite{Crespo2005}.

In conclusion  we proved numerically the dispersion managed model of mode-locking using what is to our knowledge 
the most complete 1-D simulation of an actual ultrafast Ti:Sapphire laser.
Moreover, we show how the pulse acquires and preserves a $smooth$ spectral phase in its propagation inside the cavity. These results are in very good 
agreement with experimental measurements of the spectrum outside the OC. A
detailed experimental demonstration of pulse evolution as
predicted by this model is presently under development. This work was partly supported by FCT Grant No.
POCTI/FIS/48709/2002, Portuguese Ministry of Science, co-financed by FEDER. M. V. Tognetti's e-mail address is marco.tognetti@fc.up.pt.

\newpage

\section*{List of Figure Captions}

Fig. 1 (a) Schematic diagram of the Ti:sapphire laser oscillator.
(b) Total round trip intracavity GDD. (c) Total cavity gain
$G_{tot}$ assuming a peak crystal gain of 1.04.

\bigskip
\noindent Fig. 2 Normalized intracavity spectrum (solid line) and phase (dotted line) before the OC after
$(a)\,20$, $(b)\,500$, $(c)\,1000$ round trips inside the cavity. (d)
Simulated (solid line) and measured (dashed line) final spectra
outside the OC.

\bigskip
\noindent Fig. 3 Laser cavity dispersion map. $A,B,C,D,E$ and $F$ are reference points corresponding to the spectra,
 phases, and temporal profiles reported in figure\,\ref{figure_4_tognetti_crespo} (see the text for more details).

\begin{figure}[h]
 \centerline{\scalebox{0.3}{\includegraphics[angle=270]{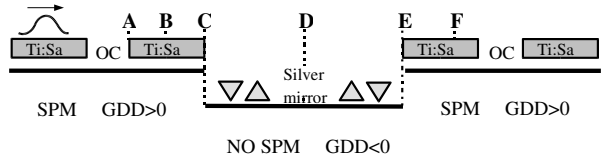}}}
\caption{Laser cavity dispersion map. $A,B,C,D,E$ and $F$ are reference points corresponding to the spectra,
 phases, and temporal profiles reported in figure\,\ref{figure_4_tognetti_crespo} (see the text for more details).}
 \label{figure_3_tognetti_crespo}
\end{figure}

\begin{figure}[h]
 \centerline{\scalebox{0.8}{\includegraphics{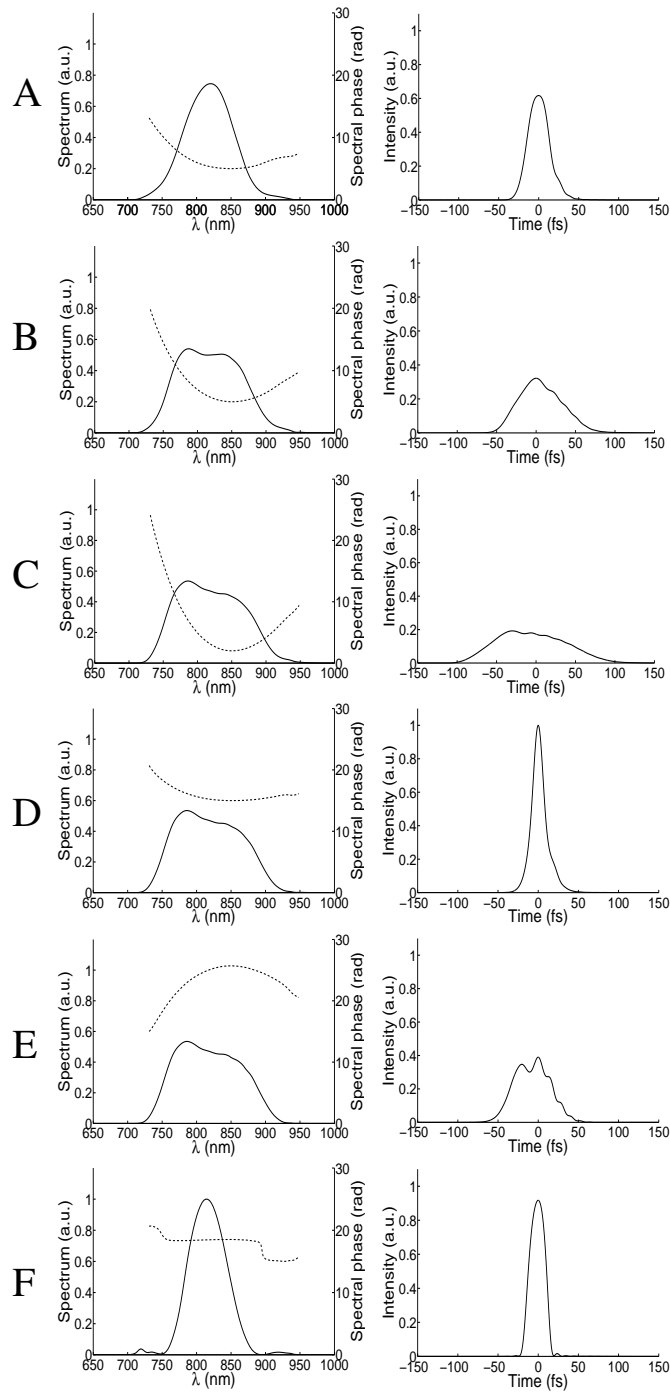}}}
\caption{Spectra, phases and temporal profiles corresponding to the reference
points reported in figure\,\ref{figure_3_tognetti_crespo}.}
 \label{figure_4_tognetti_crespo}
\end{figure}

\bigskip 
\noindent Fig. 4 Spectra, phases and temporal profiles corresponding to the reference
points reported in figure\,\ref{figure_3_tognetti_crespo}.
 
\newpage

\newpage

\newpage

\newpage

\newpage

\end{document}